\documentclass[aps,prepre,reprint,amsmath,amssymb,showpacs,superscriptaddress]{revtex4-2}
\usepackage{graphicx}
\usepackage{dcolumn}
\usepackage{bm}
\usepackage{color}
\usepackage{lipsum}
\usepackage{hyperref}
\usepackage{mathrsfs} 
\usepackage{lineno}
\usepackage{siunitx}
\usepackage{multirow}

\hypersetup{colorlinks=true,
            linkcolor=blue,
            anchorcolor=blue,
            citecolor=blue,
			urlcolor=blue}

\begin{document}

\title{Large-scale Kinetic Simulations of Colliding Plasmas within a Hohlraum of Indirect Drive Inertial Confinement Fusions}

\author{Tianyi Liang}
	\affiliation{Institute for Fusion Theory and Simulation, School of Physics, Zhejiang University, Hangzhou 310058, China}
\author{Dong Wu}
    \email{dwu.phys@sjtu.edu.cn}
    \affiliation{Key Laboratory for Laser Plasmas and School of Physics and Astronomy, and Collaborative Innovation Center of IFSA, Shanghai Jiao Tong University, Shanghai 200240, China}
\author{Xiaochuan Ning}
	\affiliation{Institute for Fusion Theory and Simulation, School of Physics, Zhejiang University, Hangzhou 310058, China}
\author{Lianqiang Shan}
	\affiliation{Science and Technology on Plasma Physics Laboratory, Research Center of Laser Fusion, CAEP, Mianyang 621900, China}
\author{Zongqiang Yuan}
	\affiliation{Science and Technology on Plasma Physics Laboratory, Research Center of Laser Fusion, CAEP, Mianyang 621900, China}
\author{Hongbo Cai}
	\affiliation{Institute of Applied Physics and Computational Mathematics, CAEP, Beijing 100094, China}
\author{Zhengmao Sheng}
	\email{zmsheng@zju.edu.cn}
	\affiliation{Institute for Fusion Theory and Simulation, School of Physics, Zhejiang University, Hangzhou 310058, China}
\author{Xiantu He}
	\affiliation{Institute for Fusion Theory and Simulation, School of Physics, Zhejiang University, Hangzhou 310058, China}
\date{\today}

\begin{abstract}
	The National Ignition Facility has recently achieved successful burning plasma and ignition using the inertial confinement fusion (ICF) approach. However, there are still many fundamental physics phenomena that are not well understood, including the kinetic processes in the hohlraum. Shan et al. [Phys. Rev. Lett, 120, 195001, 2018] utilized the energy spectra of neutrons to investigate the kinetic colliding plasma in a hohlraum of indirect drive ICF. However, due to the typical large spatial-temporal scales, this experiment could not be well simulated by using available codes at that time. Utilizing our advanced high-order implicit PIC code, LAPINS, we were able to successfully reproduce the experiment on a large scale of both spatial and temporal dimensions, in which the original computational scale was increased by approximately 7 to 8 orders of magnitude. When gold plasmas expand into deuterium plasmas, a kinetic shock is generated and propagates within deuterium plasmas. Simulations allow us to observe the entire progression of a strong shock wave, including its initial formation and steady propagation. Although both electrons and gold ions are collisional (on a small scale compared to the shock wave), deuterium ions seem to be collisionless. This is because a quasi-monoenergetic spectrum of deuterium ions can be generated by reflecting ions from the shock front, which then leads to the production of neutrons with unusual broadening due to beam-target nuclear reactions. This work displays an unprecedented kinetic analysis of an existing experiment, shedding light on the mechanisms behind shock wave formation. It also serves as a reference for benchmark simulations of upcoming new simulation codes and may be relevant for future research on mixtures and entropy increments at plasma interfaces.
\end{abstract}
\maketitle


\section{\label{sec:intro}INTRODUCTION}

	The recent experiments carried out at the National Ignition Facility (NIF) \cite{Abu2022, Zylstra2022, Hartouni2023, Zylstra2022pre, Atzeni2022, Zylstra2021prl} have successfully validated the feasibility of controllable inertial confinement fusion (ICF). This achievement is considered a noteworthy milestone in the endeavor to attain inexhaustible and environmentally friendly sources of energy. This remarkable achievement can be attributed to the collaborative efforts involving various technological advancements and physical insights. Some notable contributions include the utilization of high-density-carbon (HDC) capsules in low-gas-fill hohlraums \cite{Hopkins2018} and the implementation of the ``BigFoot'' (BF) \cite{Baker2020} scheme. Moreover, a comprehensive set of scaling laws \cite{Hurricane2019} and evaluation metrics \cite{Zylstra2021} based on the analysis of experimental data have been consolidated, serving as crucial tools in comprehending the underlying physical mechanisms of ICF and providing guidance for the development of pertinent experimental designs. However, from a theoretical standpoint, there remain numerous foundational physics concepts that are not yet comprehensively understood, particularly those that are closely tied to kinetic and non-equilibrium processes. 

	\begin{figure*}[htbp]
		\centering
		\includegraphics[scale=0.4]{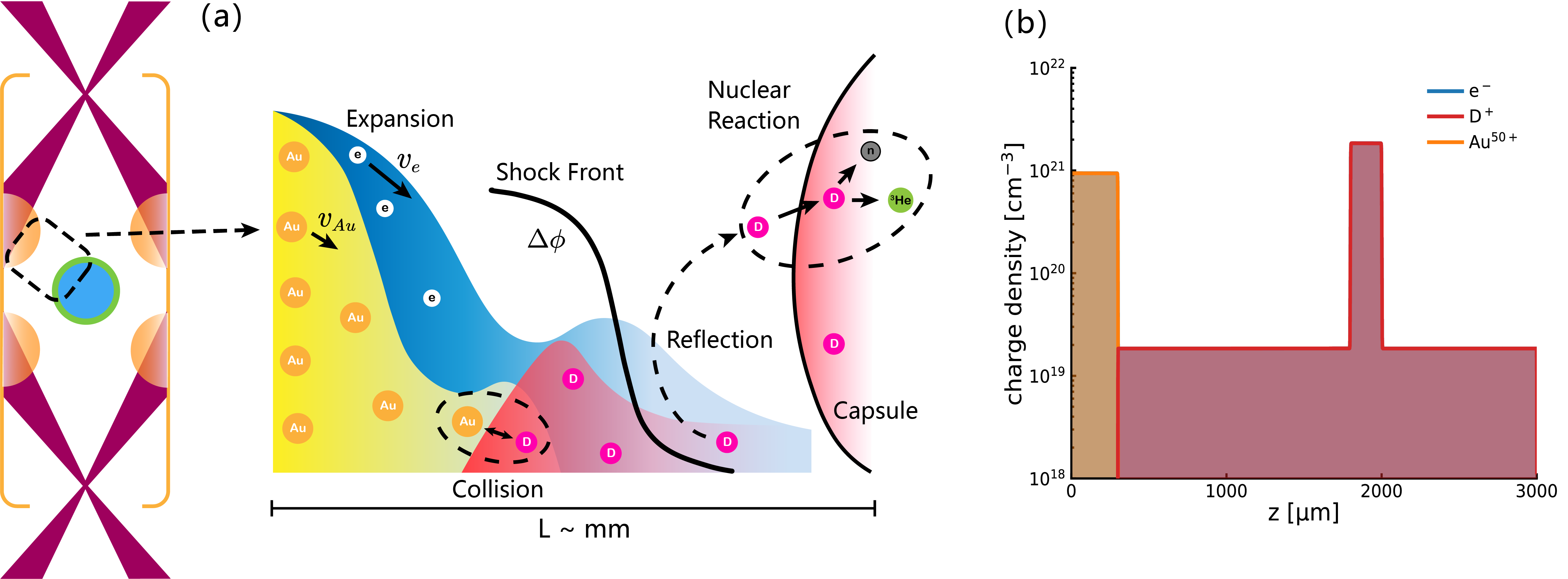}
		\caption{Subgraph (a) depicts the schematic diagram of the colliding plasmas within a hohlraum. It illustrates various phenomena such as collisional interaction, shock formation, collisionless ion reflection, and nuclear reactions. Subgraph (b) represents the simulation setup.}
		\label{fig:fig1} 
	\end{figure*}

	In vacuum or near-vacuum hohlraums \cite{Pape2016}, as depicted in Fig.~\ref{fig:fig1} (a), the expansion of high-Z plasma from the inner wall and its collision with the blow-off from the capsule or the low-density-fill gas \cite{Hopkins2015prl, Hopkins2015}, can lead to the dominance of kinetic effects. This is due to the ion-ion mean free paths being larger than the size of the interaction regions. The existence of these kinetic effects has been substantiated through experiments and simulations \cite{Shan2018, Pape2014, Li2010}, and their influence on the implosion process of ICF has been partially elucidated \cite{Rinderknecht2018}. To attain high-gain laser fusion, it is imperative to thoroughly comprehend and regulate any notable physics phenomena that may arise during ICF processes. Therefore, the study of kinetic effects and their impact on energy depositions, implosion symmetries, and other inherent plasma properties holds significant importance.

	Recently, in an experiment conducted by Shan et al. \cite{Shan2018}, a kinetic colliding plasma was observed within a hohlraum of indirect drive ICF. This observation was made by measuring the energy spectra of neutrons. The width of the spectra in this experiment is unusually large. To explain this, they conducted collisionless particle-in-cell (PIC) simulations to investigate the interactions between the gold plasma and the low-density-fill gas. It was found that the kinetic shock wave reflects upstream ions and results in beam-target fusion. This phenomenon could, in turn, explain the broadening of the neutron energy spectrum. However, the PIC simulation is somewhat ideal due to the absence of collisional effects and nuclear reactions. Moreover, the scale of the PIC simulation was only several micrometers, much smaller than the scale of real physics, which is hundreds of micrometers in the experiment.
	
	As the conventional PIC method is constrained by computational resources, making it unsuitable for large-scale simulations, scholars have put forward hybrid approaches that combine the advantages of both fluid and kinetic PIC methods \cite{Welch2001, Cai2021, Thoma2017}. While in the context of the fluid component, the authors of this study still rely on empirical coefficients, such as the flux limiter and electron-ion coupling coefficients, for their treatment. Additionally, the authors make use of various approximations in different forms. 
	The utilization of the hybrid approach assumes that the electrons must adhere to the condition of fluid approximation, wherein the mean free paths of the electrons should be smaller than the spatial resolution of the simulation. But in hohlraums or other laboratory astrophysics experiments, where the electron density ranges $10^{18}\:\rm{cm^{-3}}$ to $10^{20}\:\rm{cm^{-3}}$, the behavior of electrons cannot be accurately described using fluid dynamics. As a result, hybrid approaches may not be applicable in these scenarios. To handle a wide range of time scales, space scales, and densities, we developed a high-order implicit multidimensional PIC method. With an appropriate arrangement of space and time, the proposed method can significantly minimize numerical errors. Additionally, the utilization of a higher-order interpolation method can effectively reduce numerical noises. This reduction in numerical noise is particularly useful for simulating large-scale kinetic processes using the PIC method.

	In this paper, the experiment conducted by Shan et al. \cite{Shan2018} was simulated and analyzed using the high-order implicit PIC code LAPINS.  The simulation performed was analyzed in a large-scale set-up that closely resembles the real experiment. In our simulations, not only the density distributions and ion reflections associated with the shock wave were figured out, but also the neutron spectra that arise from the nuclear reactions between the reflected ions and the capsule were self-consistently generated. Considering the collision effects, the expansion of gold plasma manifests as a central rarefaction wave, instead of an isothermal one simulated in the reference \cite{Shan2018}. When the gold plasma expands into the deuterium plasma, it initiates the generation of a kinetic shock wave, which subsequently propagates within the deuterium plasma. The shock wave under consideration is situated in the intermediate region between collisionless and collisional regimes. The primary mechanisms accountable for dissipation in this particular context involve both the reflections of upstreaming deuterium ions and collisional processes. Deuterium ions exhibiting quasi-monoenergetic spectra can be produced through the collisionless electrostatic shock wave mechanism, resulting in the generation of neutrons with significantly broadened energy distribution due to beam-target nuclear reactions.

	The paper is organized as follows. A brief introduction to the high-order implicit PIC method and the pairwise nuclear reaction algorithm of the LAPINS code is described in Section \uppercase\expandafter{\romannumeral2}. In Section \uppercase\expandafter{\romannumeral3}, the simulation results are presented and analyzed in detail. Finally, the discussion and conclusion are displayed in Section \uppercase\expandafter{\romannumeral4}.

\section{\label{sec:algo} Simulation methods} 
 
	The LAPINS code has undergone significant advancements over the years, including the incorporation of collision effects \cite{Wu2017}, the consideration of quantum degeneracy \cite{Wu2020}, and the adoption of a hybrid-kinetic approach \cite{Wu2023}. In this section, a brief introduction to the module of nuclear reactions \cite{Wu2021}, and the high-order implicit PIC method \cite{Wu2019} utilized in our simulation.

\subsection{\label{sec:hybrid} IMPLICIT PIC MODEL}
	In the LAPINS code, a high-order implicit multidimensional PIC method has been devised to effectively address the complexities of astrophysics and dense plasmas. The spatial-temporal arrangement is established by employing Yee's algorithm in conjunction with a leapfrog algorithm to simulate the propagation of electromagnetic fields and the advancement of particles. Specifically, the charge density is positioned at the Yee cell centers, while the electric fields and current density are staggered upwards to the cell faces. Additionally, the magnetic fields are located at the cell edges. This arrangement considers the discretization of the Faraday and Ampere equations. It can be demonstrated that the equation $\nabla\times B=0$ is consistently fulfilled when it is initially valid.
	
	Our field solver algorithm efficiently tackles the numerical instabilities commonly encountered in explicit PIC methods when using relaxed time steps and grid resolution. The algorithm used in this study effectively addresses the problem of numerical cooling, a common issue in standard implicit PIC methods, by employing a pseudo-electric-field approach. The violation of Gauss's law in the cell at time $t^{n+1}$ is denoted as $F^{n+1}=\nabla\cdot E^{n+1}-\rho^{n+1}$. It is used to calculate the additional pseudo electric field $E^{n+1}_\text{psd}=d\Delta t \nabla F^{n+1}$, with the introduction of a customized dimensionless number $d$. Consequently, in each time increment, the overall electric field $E^{n+1}$ exerted on particles needs to be updated by adding the pseudo electric field, which can be expressed as $E^{n+1}=E^{n+1}+E^{n+1}_\text{psd}$. 
	
	The particle pusher algorithm is a combination of the standard Boris particle pusher and the Newton-Krylov iteration method. This algorithm has the potential to greatly enhance precision accuracy, surpassing the standard Boris particle pusher by several orders of magnitude. Additionally, it offers a substantial reduction in iteration time compared to the pure Newton-Krylov method. For further information, readers are encouraged to consult our recent publication \cite{Wu2019}.

\subsection{\label{sec:nuclear} MODEL OF NUCLEAR REACTION}

	We have successfully developed a model for pairwise nuclear reactions involving weighted particles at relativistic energies \cite{Wu2021}. It is worth mentioning that the particle-pairing routine used for fusion reactions is identical to the one employed for binary collisions, known as the Takizuka-Abe algorithm. In every spatial cell, pairs of particles engaging in nuclear reactions are randomly selected. Energy and momentum exchanges are calculated for every pair of particles and performed in the center-of-momentum (CM) frame. Our model is additionally applicable in the context of the relativistic regime. In the CM frame, the probability of the reaction, denoted as $P_\text{ab}$, is determined by the calculation of 
	\begin{equation}
		P_\text{ab} = n_\text{min}\sigma \rm{v}_{rel,CM}\gamma_\text{CM}\Delta t, 
		\label{eq:eq1}
	\end{equation}
	where $n_\text{min}$ is the minimum density between particle species $a$ and $b$, $\sigma_\text{ab}$ denotes the cross-section of nuclear reaction, and $\Delta t$ is the simulation time step, which is multiplied by a factor of $\gamma_\text{CM}$ when taken into account in the CM frame. The nuclear reaction cross sections used in the program are obtained from the International Atomic Energy Agency (IAEA) and are stored within the code as Legendre polynomial coefficients. To calculate these values, the relative velocity between the two particles in the CM frame is required and provided by
	\begin{equation}
		\rm{v}_\text{rel,CM} =\left|\frac{\mathbf{v}_{a,CM}-\mathbf{v}_{b,CM}}{1-\mathbf{v}_{a,CM}\cdot\mathbf{v}_{b,CM}} \right|. 
		\label{eq:eq2}
	\end{equation}
	Therefore, the yield of nuclear reaction for each pair of macro-particles is determined as
	\begin{equation}
		Y_\text{ab}=w_\text{min}P_\text{ab},
		\label{eq:eq3}
	\end{equation}
	where $w_\text{min}$ represents the minimum weight between macro-particles $a$ and $b$. To enhance the calculation accuracy, we have introduced a variable parameter $F_\text{multi}$ to increase the probability of the reaction while decreasing the weight of the products, denoted as $w_\text{min}\to w_\text{min}/F_\text{multi}$ and $P_\text{ab} \to P_\text{ab}F_\text{multi}$, while keeping $Y_\text{ab}$ unaltered. For details of the nuclear fusion scheme, one can refer to our recent paper and other relevant publications \cite{Wu2021}.
 
\section{\label{sec:simu} SIMULATION RESULTS}

	\begin{figure*}[htbp]
		\centering
		\includegraphics[scale=0.6]{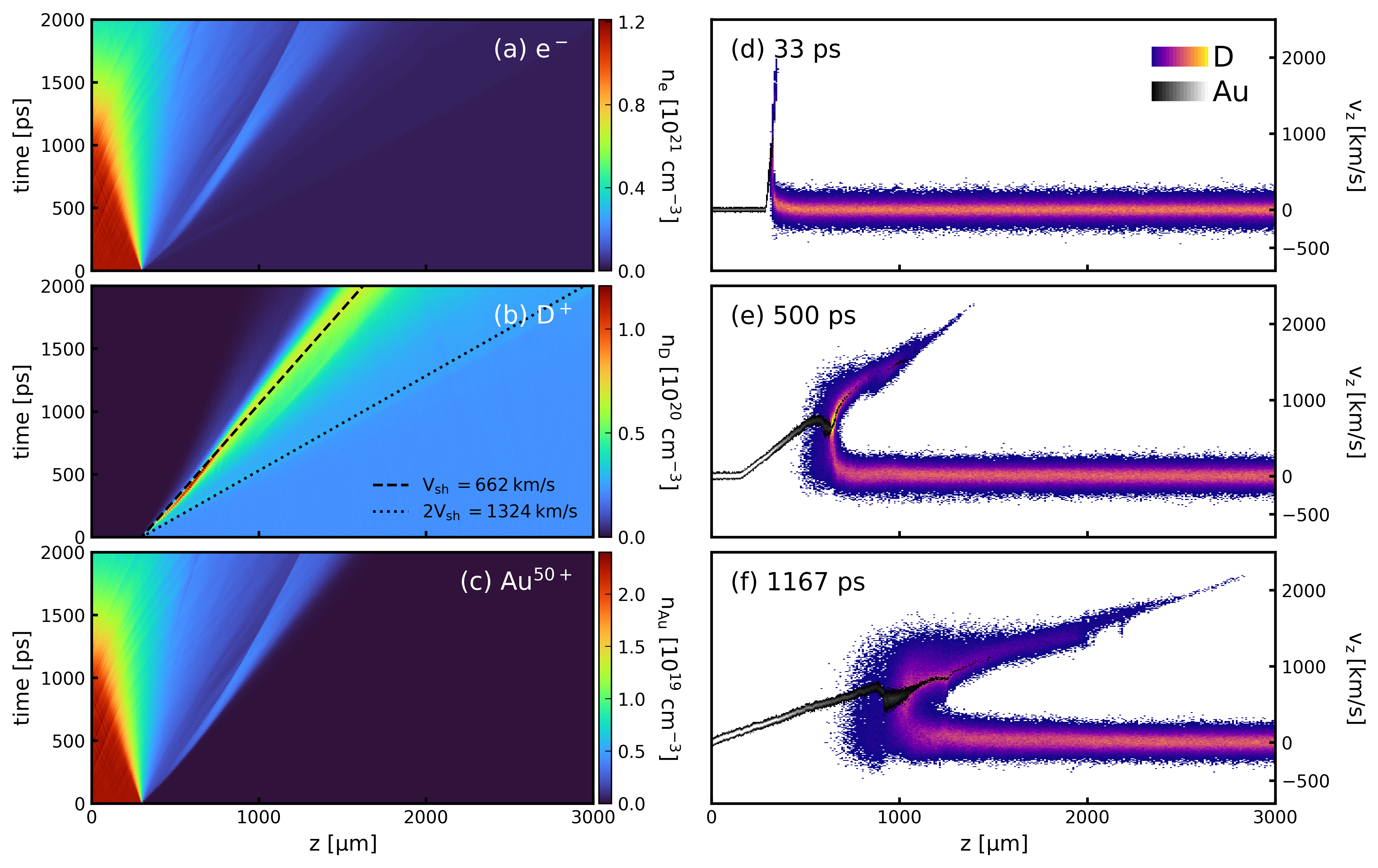}
		\caption{The subgraphs (a)-(c) depicted in the left part illustrate the temporal evolution of the density profiles of electrons, deuterium ions ($\rm{D^+}$), and gold ions ($\rm{Au^{50+}}$), respectively. The black dashed line depicted the trajectory of the shock front, characterized by a velocity of $V_{sh}=662\:\rm{km/s}$, while the dotted line represents the trajectory with double the velocity, $2V_{sh}$. The subgraphs (d)-(f) on the right side depict the phase space distributions of gold ions and deuterium ions at different times. The white-black colormap and colorful colormap are used to represent the phase density of gold ions and deuterium ions, respectively.}
		\label{fig:fig2} 
	\end{figure*}

	To conduct a more in-depth analysis of the experimental data \cite{Shan2018}, the LAPINS code is employed, utilizing parameters that closely resemble those of the experimental setup. This simulation involves the collision interaction between the ablated gold plasma and deuterium gas, as depicted in Fig.~\ref{fig:fig1} (b). The simulation box has a length of $L=3\:\rm{mm}$, which is resolved by $3000$ cells. Each cell contains $1000$ macro-particles for electrons and $400$ macro-particles for various ions. Simulation times in this study are on the order of nanoseconds, suggesting that the scale of the simulation is approximately 7 to 8 orders of magnitude larger than that of traditional PIC codes, which typically operate on a temporal scale of picoseconds and a spatial scale of micrometers. 
	
	The gold plasma is located on the left side of the simulation box. The ionization degree of gold ions is fixed at $Z=50$, while the initial ion temperature is $T_\text{Au}=100\:\rm{eV}$. The initial electron temperature of the gold plasma is $T_{e1}=3000\:\rm{eV}$, while the electron density is $n_{e1}=Zn_\text{Au}=1.0 \times 10^{21}\:\rm{cm^{-3}}$. Full-ionized deuterium plasma is situated on the right side of the simulation box, with a density of $n_{e0}=n_\text{D}=2.0\times 10^{19}\:\rm{cm^{-3}}$ and temperature of $T_\text{D}=T_\text{e0}=100 \: \rm{eV}$. Electrostatic shock structure is generated from the expansion of gold plasma into deuterium plasma. The gold plasma is commonly recognized as the downstream region located behind the shock, whereas the deuterium plasma is considered the upstream region situated ahead of the shock. An absorbing layer is designated to function as the capsule characterized by a high density of $2.0\times 10^{21}\:\rm{cm^{-3}}$ and a low temperature of $1\:\rm{eV}$. The speed of sound in deuterium plasma is $c_s = \sqrt{{kT_{e0}}/{m_\text{D}}}=126.4\:\rm{km/s}$. It is important to note that there is no initial drift velocity for any species in the plasmas.   
	
	In the initial stages of the expansion, the hotter electrons move faster than the gold ions. This differential motion generates an electrostatic field characterized by charge separation, commonly referred to as the sheath electrostatic field. This electric field plays an important role in both the expansion of the gold plasma and the reflection of the deuterium plasma. Plasma expansion is a phenomenon wherein the internal energy of plasma is converted into kinetic energy. When considering the collisional effect, the expansion behaves as a centered rarefaction wave \cite{Atzeni2004} instead of an isothermal one, as the deuterium plasma acts as a piston. During the process of expansion, the velocity of the gold plasma exhibits a linear increase until it reaches its maximum value. It is illustrated in Fig.~\ref{fig:fig2} (d)-(f), which depict the phase space distributions of gold ions and deuterium ions at various time intervals. The collision frequency between deuterium and gold ions, $\nu_{\text{D,Au}} = 4/3\pi^{1/2}({Z_\text{Au}Z_\text{D}}/{4\pi\varepsilon_0})^2n_\text{D}{m_\text{D}^{-1/2}T_\text{D}^{-3/2}}\ln\Lambda$, is estimated to be in the range of $10^{12} \sim 10^{13}/\rm{s}$, and the mean free path is approximately $l_{\text{D,Au}}\approx 0.15\:\rm{\mu m}$. In the context of the penetration region, the presence of intense collisions between deuterium ions and gold ions hinders the entry of deuterium ions into the downstream region. 
	
	Therefore, during the initial formation of the shock wave, the shock strength denoted as $\delta = n_\text{Dmax}/n_\text{D}$ and referred to as the density compression ratio, rapidly reaches a value of 6. Subsequently, after approximately 1 ns of simulation time, it gradually decreases to 3.5 and remains stable. The decrease can be attributed to the expansion of the dense deuterium plasma in the shock region to the upstream deuterium plasma. In the stable stages, the shock strength $\delta$ remains below the upper limit determined by the Rankine-Hugoniot equations, indicating that $\delta$ is less than 4. 
	
	\begin{figure}[htbp]
		\centering
		\includegraphics[scale=0.45]{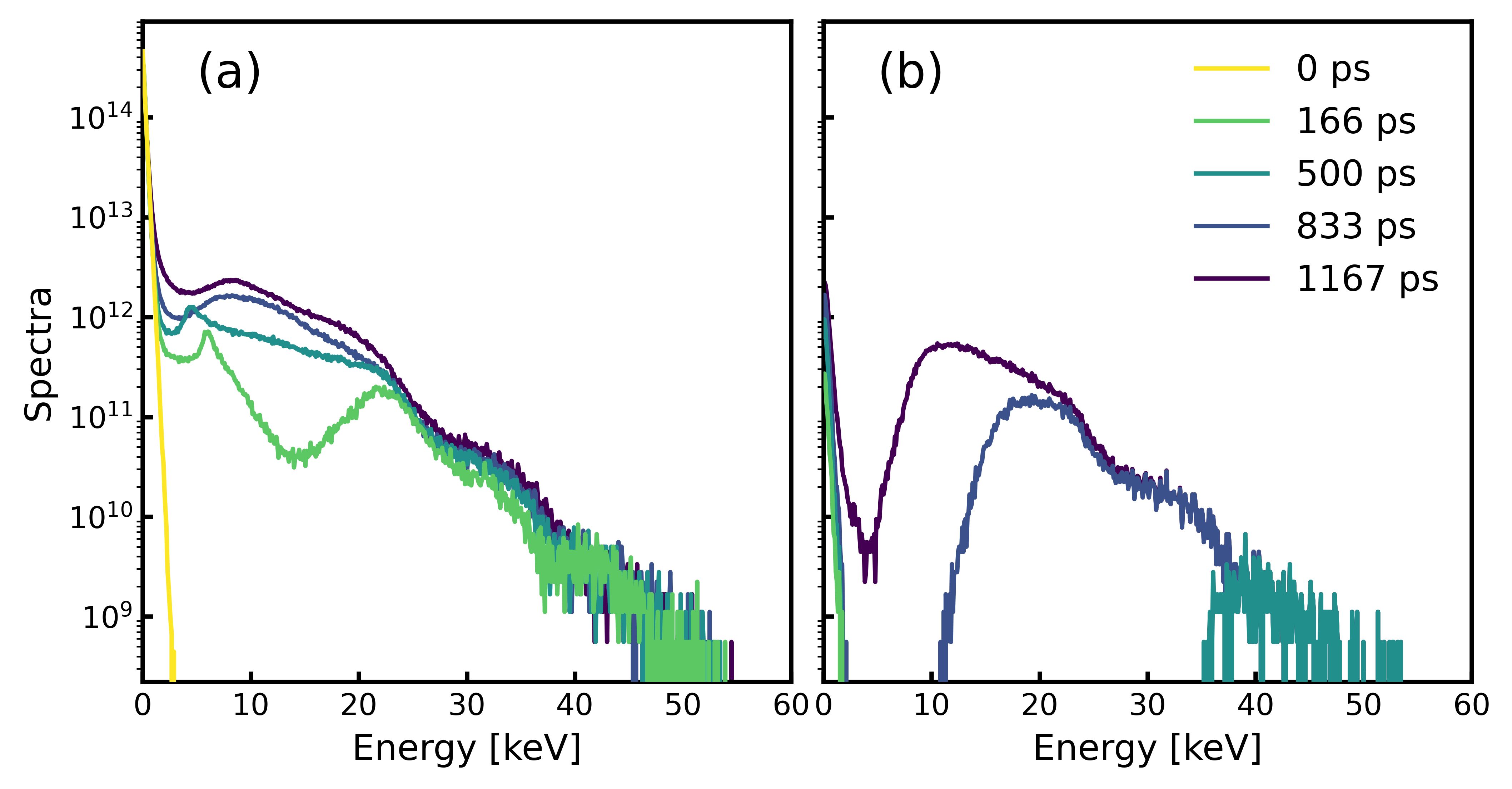}
		\caption{The temporal evolution of the energy spectra of deuterium ions. The total spatial energy spectra are presented in subgraph (a), while the time-integrated energy spectra obtained at the location of $z=1200\:\rm{\mu m}$ are displayed in subgraph (b).}
		\label{fig:fig3}
	\end{figure}

	One of the notable kinetic properties exhibited by electrostatic shock is the reflection of upstream ions. The electric potential difference, $\Delta\phi$, can reflect any ions with kinetic energies lower than it, as expressed by $m_iv_i^2/2<e\Delta\phi$ \cite{Sagdeev1991, Balogh2013}. In our simulations, it is evident that the reflection of deuterium ions can be categorized into two distinct phases. Firstly, there is a pronounced acceleration of the sheath electric field at the initial stage, which contributes to the high-energy portion of the reflected ions. Secondly, there is a consistent reflection of the electrostatic shock wave, with a speed that can reach up to twice the speed of the shock wave ($2V_\text{sh}$) in the laboratory frame. The velocity of the shock, denoted as $V_\text{sh}$ was determined based on the temporal evolution in the density distribution of electrons and ions depicted in Fig.~\ref{fig:fig2} (a)-(c). The measured velocity of the shock wave is $V_\text{sh}=662 \: \rm{km/s}$ and the Mach number is $M=V_\text{sh}/c_s=5.25$. 
	
	In the phase space depicted in Fig.~\ref{fig:fig2} (d)-(f), the reflected deuterium ions are accelerated to velocities exceeding $2000\:\rm{km/s}$, with a steady velocity of approximately $1300\:\rm{km/s}$, which is nearly twice $V_\text{sh}$. In the theoretical framework of the KdV-B equation, the phenomenon of steady reflection can be understood as a form of dissipation. The collisional effect does not provide sufficient dissipation due to the low density of the deuterium plasma. As a result, reflection compensates for dissipation, which plays an important role in forming the shock as well as dispersion and nonlinearity. When dissipation, dispersion, and nonlinearity are balanced, the shock wave can propagate steadily, otherwise, it will collapse or disappear \cite{Balogh2013}.  
	
	Diagnosing the evolution of the total energy spectra of deuterium ions over time, as depicted in Fig.~\ref{fig:fig3} (a), it is evident that the energetic ions generated through reflection exhibit substantial deviations from the initial narrow Maxwell distribution. A time-integral diagnostic plane is established at the location of $z=1200\:\rm{\mu m}$ to record the energy spectra of reflected ions, which can arrive at the absorbing layer, as shown in Fig.~\ref{fig:fig3} (b). Deuterium ions exhibiting quasi-monoenergetic spectra are observed, reaching a maximum energy of approximately $50\:\rm{keV}$, with a full width at half maximum (FWHM) $\Delta E_\text{D} = 30\:\rm{keV}$. The spectra undergo broadening when the shock wave passes through the diagnostic plane. When these reflected ions are deposited inside the capsule, significant low-mode asymmetry may occur in the implosion process \cite{Glenzer2010}, as well as the beam-target nuclear reactions. 
	
\begin{figure}[htbp]
		\centering
		\includegraphics[scale=0.28]{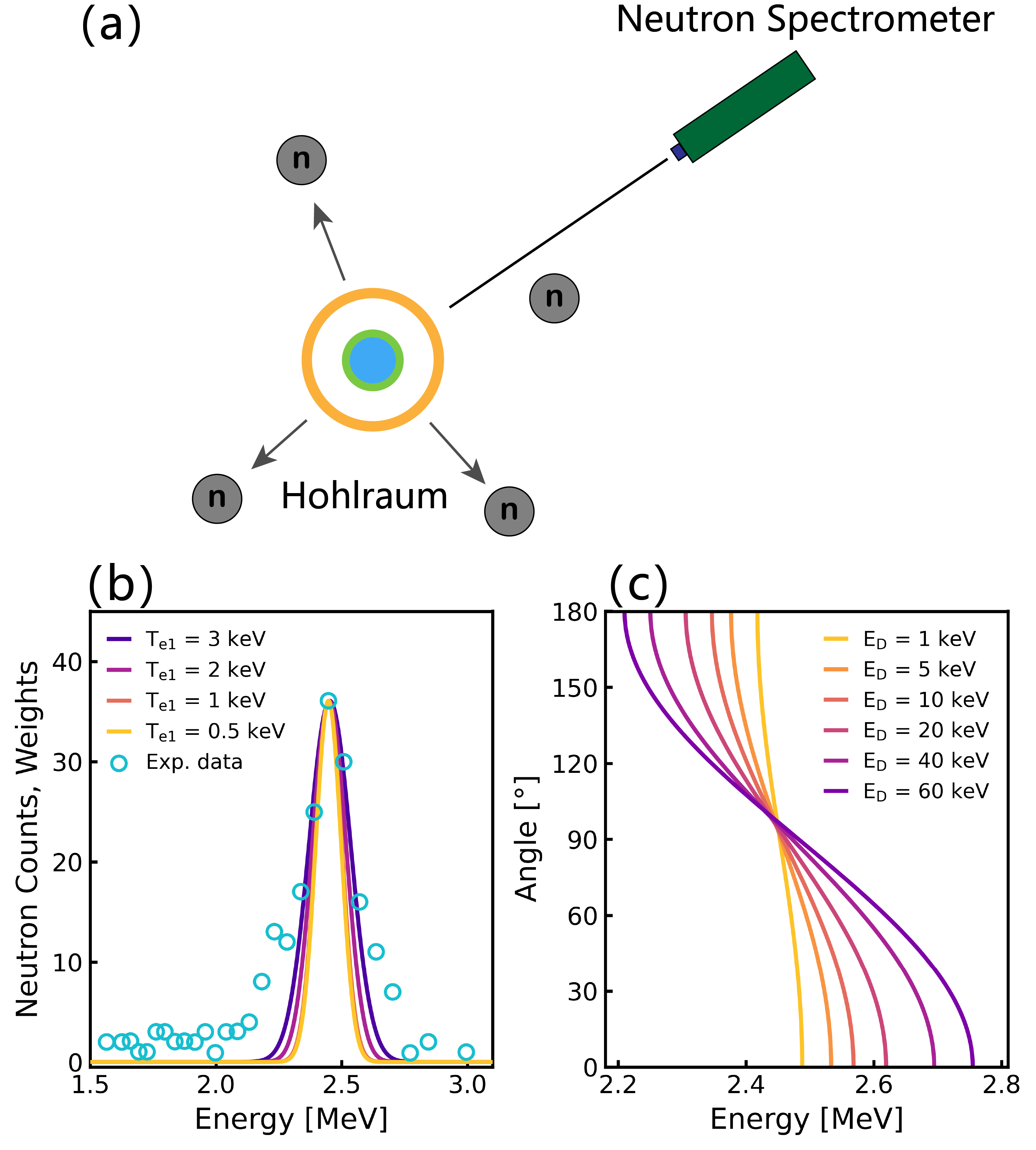}
		\caption{Subgraph (a) is a top-view schematic diagram of the setup of a neutron spectrometer in a hohlraum experiment. Subgraph (b) represents the comparison of the neutron spectra obtained in the simulation with that of the experimental data. The solid lines of different colors are the simulated data over different electron temperatures $T_{e1}$, fitted with Gaussian profiles. The experimental data presented by Shan et al. \cite{Shan2018} is depicted as light blue circles. Subgraph (c) represents the numerical results obtained from Eq.~\ref{eq:eq4} for various incident energies of deuterium ions, denoted as $E_\text{D}$.}
		\label{fig:fig4}
	\end{figure}

	To simulate the deposition of reflected ions inside the capsule, an absorbing layer located at $2000\:\rm{\mu m}$ is set to capture the energetic ions accelerated by the shock wave. As the probability of the nuclear reaction $\rm{D(D,n)^3He}$ is very small for reflected ions with a speed of several hundred kilometers per second and a temperature of $1\:\rm{eV}$ in the plasma layer, we have set the parameter $F_{\text{multi}}$ as $10000$ using more than 1 million deuterium macro-particles. The statistical and fitting data of the neutron spectra are shown in Fig.~\ref{fig:fig4} (b). The peak energy is $2.45\:\rm{MeV}$ with a FWHM of $\Delta E_{n}=0.3\:\rm{MeV}$. The relationship between FWHM and $T_\text{D}$ in thermonuclear conditions can be described as $\Delta E_{n} = 2\sqrt{\ln{2}\langle E_n\rangle T_\text{D}}= 82.5T_\text{D}^{1/2}$ \cite{Brysk1973}, where $T_\text{D}$ is in keV and $\langle E_n \rangle = 2.45\: \rm{MeV}$. If we assume that this unusual broadening is due to thermonuclear fusion, the estimated temperature ($T_\text{D}$) is approximately $13 \:\rm{keV}$, which is significantly higher than the temperature achievable for the high-density absorbing layer ($1\:\rm{eV}$). 
	
	By utilizing double-differential cross-sections, $\rm{d}^2\sigma/\rm{d}E\rm{d}\Omega$, the energies of emitted neutrons can be observed as a function of scattering angle $\theta$ in the CM frame, denoted as,
	\begin{align}
		& E_n = \frac{E_\text{D} m_n}{4m_\text{D}} \left[\frac{2m_\text{D} - m_{n}}{m_n}\left(2\frac{Q}{E_\text{D}}+1\right)  \right. \nonumber \\
		& \left. + 2\sqrt{\frac{2m_\text{D} - m_{n}}{m_{n}} \left(2\frac{Q}{E_\text{D}}+1\right) } \cos{\theta} + 1 \right],
		\label{eq:eq4}
	\end{align}
	where $E_\text{D}$ is the incident energy of deuterium ions, and $Q=3.27\:\rm MeV$ is the released energy of $\rm{D(D,n)^3He}$ nuclear reaction. The numerical results obtained from Eq.~\ref{eq:eq4} for various incident energies of deuterium ions are shown in Fig.~\ref{fig:fig4} (c). When using the value of our simulations, e.g. $E_\text{D}=50\:\rm{keV}$, we find that $\Delta E_{n,\theta} = E_n(\theta=0)-E_n(\theta=\pi) = 497\:\rm{keV} $, which is well consistent with the spectra, as shown in Fig.~\ref{fig:fig4} (b). The spectra exhibit natural broadening due to beam-target reactions when the neutrons are counted by integrating the full solid angle. This approach was used in our simulation. As illustrated in Fig.~\ref{fig:fig4} (a), during the experiment, the neutron spectrometer was arranged in a narrow solid angle configuration to capture neutrons with varying scattering angles $\theta$ emitted from the entire hohlraum. This experimental setup is mathematically equivalent to our one-dimensional simulations, where all neutrons are accounted for by integrating over the full solid angle. When comparing the FWHM of the neutron spectra in our simulations with that of the experiment data (represented by light blue circles in Fig.~\ref{fig:fig4} (b)), a high level of consistency is observed. 

\section{\label{sec: disc}DISCUSSION AND CONCLUSION}
	
\begin{figure}[htbp]
		\centering
		\includegraphics[scale=0.45]{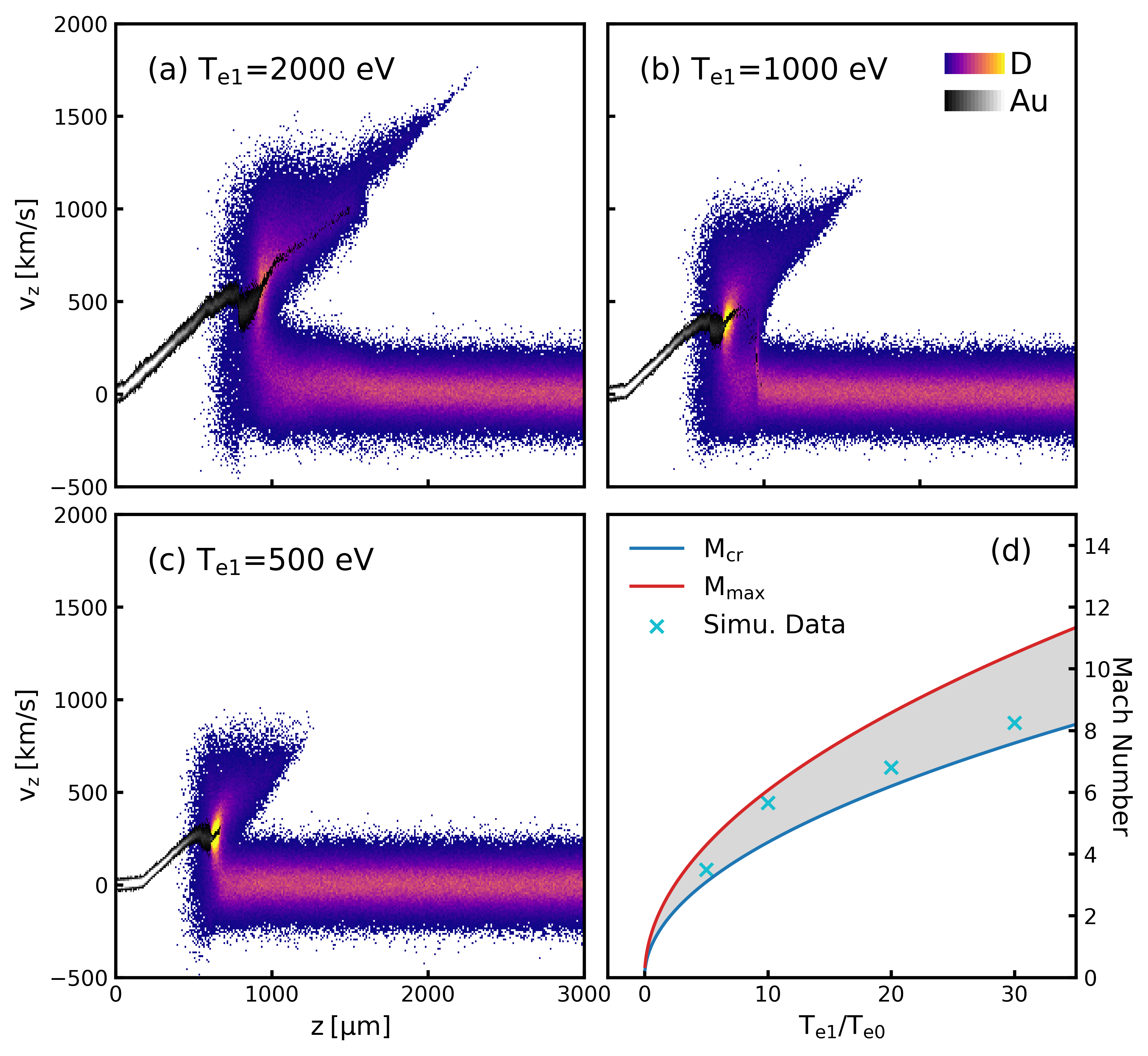}
		\caption{Subgraphs (a)-(c) depict the phase space distributions for gold ions and deuterium ions concerning electron temperature $T_{e1}$ of the gold plasma. The white-black colormap and colorful colormap represent the phase density of gold ions and deuterium ions, respectively. Subgraph (d) depicts a comparison of Mach numbers obtained from simulations (represented by blue crosses) and theoretical calculation results from Eq.~\ref{eq:eq5} and Eq.~\ref{eq:eq6} (represented by blue and red solid lines).}
		\label{fig:fig5}
	\end{figure}

	In a hohlraum experiment, the determination of the state of the gold plasma poses a significant challenge, as it directly influences the properties of the shock wave and the behavior of the reflected ions. To elucidate the impacts, we manipulated the electron temperature $T_{e1}$ of the gold plasma. The phase space distributions are shown in Fig.~\ref{fig:fig5} (a)-(c). As $T_{e1}$ decreases, the ions accelerated by the sheath electrostatic field have lower energy and constitute a smaller portion of the reflected ions. The velocity of the shock wave, $V_\text{sh}$ also decreases. According to the collisionless laminar shock wave kinetic theory \cite{Sorasio2006}, a shock wave can arise from the collision between two plasma slabs with different temperatures and densities. Ion reflection occurs when the electrostatic potential surpasses the kinetic energy of upstream ions. There exists a critical Mach numbers $M_\text{cr}$ for reflecting ions, that can be calculated numerically by \cite{Fiuza2013},
	\begin{align}
		& M^2_\text{cr} = \frac{1}{1+\Gamma} \left[\frac{\sqrt{2}M_\text{cr}}{\sqrt{\pi}} + e^{\frac{M^2_\text{cr}}{2}}\text{Erfc}\frac{M_\text{cr}}{\sqrt{2}} - 1\right. +\Gamma\Theta \times \nonumber \\
		& \left. \left(\frac{\sqrt{2}M_\text{cr}}{\sqrt{\pi\Theta}}+ e^{\frac{M^2_\text{cr}}{2\Theta}}\text{Erfc}\frac{M_\text{cr}}{\sqrt{2\Theta}}+\frac{4M^3_\text{cr}}{3\sqrt{2\pi\Theta^3}}-1\right)\right],
		\label{eq:eq5}
	\end{align}
	where $\Gamma=n_{e1}/n_{e0}$, and $\Theta=T_{e1}/T_{e0}$, are the ratios of upstream and downstream electron densities and temperatures. In the simulation, the $\Gamma=51$ is fixed. When $M^2 \gg 1$ and $M^2 \gg \sqrt{\Theta}$, there is an upper limit to the Mach number \cite{Sorasio2006}, 
	\begin{equation}
		M_\text{max} \simeq \frac{3(\Gamma+1)}{\Gamma} \sqrt{\frac{\pi\Theta}{8}}.
		\label{eq:eq6}
	\end{equation}
	The Mach numbers estimated in our simulations via different $\Theta$ are presented in Fig.~\ref{fig:fig5} (d), which is well located between two lines. The Mach number can be mathematically represented as a function of $\Theta$ and $\Gamma$. Consequently, the characteristics of the shock wave are primarily determined by the distinct properties of electrons in the upstream and downstream regions.
	The corresponding neutron spectra are shown in Fig.~\ref{fig:fig4} (b). As the energy of the reflected ions decreases, the neutron energy spectra become narrower, aligning with the theoretical prediction of the beam-target reaction. The FWHM of the neutron spectra obtained from the experiments can serve as an indicator for estimating the electron temperature of the gold plasmas.

	To summarize, Shan et al. \cite{Shan2018} conducted a study on a kinetic colliding plasma within a hohlraum of indirect drive ICF by measuring the energy spectra of neutrons. However, due to the typical large spatial-temporal scales involved, this experiment could not be accurately simulated using available codes at that time. The experiment was successfully replicated at large spatial and temporal scales using our high-order implicit PIC code, LAPINS. The computational scale of our simulations is approximately 7 to 8 orders of magnitude greater than that of traditional PIC codes. When gold plasmas expand into deuterium plasmas, a kinetic shock is generated and propagates within deuterium plasmas. Through the utilization of simulations, we can observe the complete progression of a strong shock wave, encompassing its initial formation and subsequent steady propagation. Although both electrons and gold ions are collisional, deuterium ions appear to be collisionless. The quasi-monoenergetic spectra of deuterium ions can be generated by reflecting ions from the shock front. This process leads to the production of neutrons with unusual broadening due to beam-target nuclear reactions. This study provides an unprecedented kinetic analysis of an existing experiment, which contributes to our understanding of the mechanisms underlying the formation of shock waves. It can be relevant for future research on mixtures and entropy increments at plasma interfaces.

\begin{acknowledgments}
	This work is supported by National Natural Science Foundation of China (Grants No. 12075204 and No. 11875235), Science and Technology on Plasma Physics Laboratory Foundation of Chinese Academy of Engineering Physics, the Strategic Priority Research Program of Chinese Academy of Sciences (XDA250050500), and Shanghai Municipal Science and Technology Key Project (No. 22JC1401500). Dong Wu thanks the sponsorship from Yangyang Development Fund.
\end{acknowledgments}
 
\nocite{*}

\bibliography{references}

\end{document}